\documentclass[seceq,twocolumn]{jpsj3}

\title{Study of Kosterlitz-Thouless transition of Bose systems governed by a random potential using quantum Monte Carlo simulations}

\author{Hiroki KUROYANAGI, Mitsuaki TSUKAMOTO, and Makoto TSUBOTA \thanks{E-mail address: tsubota@sci.osaka-cu.ac.jp} %\\
}
\inst{Department of Physics, Osaka City University, Sumiyoshi-ku, Osaka 558-8585, Japan%\\
}
\abst{We perform quantum Monte Carlo simulations to study the 2D hard-core Bose-Hubbard model in a random potential. Our motivation is to investigate the effects of randomness on the Kosterlitz-Thouless (KT) transition and its quantum critical phenomenon. The chemical potential is assumed to be random, by site, with a Gaussian distribution. The KT transition is confirmed by a finite-size analysis of the superfluid density and the power-law decay of the correlation function. By changing the variance of the Gaussian distribution, we find that the transition temperature decreases as the variance increases. We obtain the phase diagram showing the superfluid and the disordered phases, and estimate the quantum critical point (QCP) and the critical exponent. Our results on the ground state show the existence of the Bose glass phase. Finally, we discuss what the variance of the QCP indicates from the viewpoint of $\textit{quantum percolation}$.}

\kword{Kosterlitz-Thouless transition, superfluidity, quantum Monte Carlo simulation, quantum phase transition, disordered potential}

\begin{document}
\maketitle

\section{Introduction}
 The Kosterlitz-Thouless (KT) transition\cite{Kosterlitz} is an important concept underlying two-dimensional (2D) physics. The XY model is a typical model describing the KT transition. This model describes well not only 2D XY magnets and 2D crystals but also superfluidity in thin films, which is a particularly good example to use when studying the KT transition. One of the characteristics of the KT transition can be seen in  superfluid (SF) density. Kosterlitz and Thouless theoretically predicted that the SF density $\rho_s$ shows a discontinuous jump at the transition temperature $T_{\rm KT}$ given by the normalized form
\begin{equation}
\rho_s= \frac{2}{\pi} T_{\rm KT}.
\label{SF1}
\end{equation}
%\begin{equation}
%\rho_s= \frac{2}{\pi}.
%\end{equation}
Equation~(\ref{SF1}), which is known as the universal jump, provides an important proof of the KT transition. Many experiments using a torsional oscillator\cite{Reppy,Agnolet} have observed this universal jump. Another characteristic is the correlation function of the order parameters. According to the KT theory, the phase transition results from unbinding vortex pairs. When the temperature $\textit{T}$ is lower than the transition temperature $T_{\rm KT}$, the correlation function of the order parameters shows a power-law decay as a function of the distance $\textit{r}$, $\langle \psi(x) \psi(x+r) \rangle$$\propto r^{-\eta(T)}$, which is called the quasi long-range order (QLRO). This state is stable because the excitation by vortex pairs does not destroy the correlation of the order parameters. As $\textit{T}$ increases above $T_{\rm KT}$, vortex pairs become unbound as free vortices. These free vortices destroy the QLRO, forcing the system into the disordered phase, where the correlation decays exponentially.

Finite-size effects must be taken into account in numerical simulations of the KT transition. Although finite-size scaling is the best analysis method for studying thermodynamic limits with finite-size numerical data, ordinary finite-size scaling cannot be applied to numerical simulations of the KT transition\cite{difficult} because the logarithmic correction of the system size makes it difficult to obtain reliable estimates of $T_{\rm KT}$. Several numerical studies\cite{Weber, Kawashima} have used the XY model to analyze the nature of the KT transition. These previous studies focused on the SF density to obtain $T_{\rm KT}$ . Weber and Minnhagen performed classical Monte Carlo simulations for the classical 2D XY model and used the KT renormalization group equation to analyze the numerical data instead of using ordinary finite-size analysis\cite{Weber}. They systematically studied the size dependence of the SF density at the transition temperature to reveal from the logarithmic dependence that the KT transition actually occurs in the 2D classical XY model. Harada and Kawashima verified the KT transition of the 2D quantum XY model using quantum Monte Carlo (QMC) simulations\cite{Kawashima}. They also calculated the SF density precisely and confirmed a close agreement between their numerical data and the logarithmic scaling of the KT theory. 

For about two decades, the properties of disordered Bose systems have been an interesting topics. Experimentally, these systems are relevant to $^4$He in nano-porous materials (eg. Vycor\cite{Vycor}, Aerogel\cite{Aerogel} and Gelsil\cite{Shirahama}) and the atomic Bose gas in the optical lattices\cite{Lye,White}. Theoretically, the Bose--Hubbard (BH) model gives a good description of the physics of interacting bosons,
\begin{equation}
{\hat{\cal{H}}}=-J\sum_{<i,j>}(\hat b_i^{\dagger} \hat b_j+\hat b_i \hat b_j^{\dagger}) + \frac{U}{2} \sum_i \hat n_i{(\hat n_i -1)} + \sum_i \mu_i \hat n_i,
\label{BH}
\end{equation}
where $\hat b_i^{\dagger} (\hat b_i)$ is a boson creation (annihilation) operator, $\hat n_i=\hat b_i^{\dagger}\hat b_i$, $\textit{J}$ is the hopping parameter, $\textit{U}$ is the on-site repulsive interaction, and $\mu_i$ is the random chemical potential at  site $\textit{i}$. This model corresponds to the XY model when we impose the hard-core limit ($\textit{U}\rightarrow\infty$). The hopping parameter $J$, the interaction of particles $U$, and the disorder of the chemical potential $\mu_i$ compete to determine the ground state of Bose systems. In the first attractive work\cite{Fisher}, Fisher ${et}$ ${al.}$ predicted that there are three phases at $\textit{T}=0$ in disordered Bose systems: SF phase, insulator phase, and Bose-glass (BG) phase. In the pure system, the quantum phase transition between the SF (no gap and finite compressibility) and the insulator (gapped and no compressibility) phases occurs. On the other hand, in the disordered system, there appears the intermediate localized-insulating state which is called the BG (no gap and finite compressibility) phase, in addition to the SF and the insulator phases. After the prediction of the BG phase, interests of many theoretical and numerical studies concentrate on the exsitence of the BG phase and the role of weak disorder in the vicinity of the SF-insulator critical point. While Fisher ${et}$ ${al.}$ suggested that a direct SF-insulator transition is not allowed in principle in ref.\citen{Fisher}. The direct SF-insulator transition was shown by several numerical transition\cite{Nandini1,Nandini2,Kisker}, which were actually incomplete. Recently, by overcoming the difficulty of the previous works, the strong evidence for the absence of the direct transition for weak disorder was reported\cite{glass} so that the discrepancy was fixed.

The exsistence of the BG phase in 2D disordered systems are also reported in many literature, but there is an ambiguous problem between the SF-BG phases. According to the scaling theory\cite{Fisher}, the compressibility shows the scaling form at the critical point described as
\begin{equation}
\kappa \sim \delta^{\nu(D-z)}.
\label{scaling}
\end{equation}
Here, $\delta$ is the distance to the critical point in terms of controllable parameters, and $z$ and $\nu$ are the dynamical and the correlation length exponent. Since the SF and the BG phases are compressible, the compressibility should be finite also at the critical point. Thus, we can expect that $z=2$ if the dimension $\textit{D}$ is 2. There are contributions to estimate $\textit{z}$ and $\nu$ using Monte calro simulation\cite{MC1,MC2,2D}, the dual theory\cite{dual} and the real-space renormalization group theory\cite{real}. Actually, the estimations of the value of $\textit{z}$ show the good agreement on the expectation of $\textit{z}=2$ ($\textit{z}$=2\cite{MC1,MC2,2D}, 1.93\cite{dual} and 1.7\cite{real}), but there is no good agreement between the estimated values of $\nu$ ($\nu$=0.9\cite{MC1}, 1.15\cite{MC2}, 1.38\cite{dual} and 1.4\cite{real}).

This paper describes numerical simulations of 2D disordered Bose system. Our aim in this paper is to investigate the relationship between the KT transition and the randomness and its quantum critical phenomenon. We first make the relationship clear by taking accout of the logarithmic correction due to the finite-size effects. In addition, through determination of the quantum critical point (QCP), we succeeded in estimating the critical exponent precisely. The following is an outline of this paper. In Section 2, we describe the model and the numerical method. We perform QMC simulations to study the disordered BH model for a 2D square lattice with a periodic boundary condition. The chemical potential differs from site to site according to a Gaussian distribution. In Section 3, calculations and results are shown. The KT transition is confirmed from the finite-size scaling of the SF density and the power-law decay of the correlation function. To investigate the effects of randomness, we change the variance of the distribution. Although the transition temperature decreases as the variance of the distribution increases, the KT transition still survives. We determine the critical exponent and the QCP. By calculating the SF density and the compressibility, the nature of the ground state in this disordered system is also investigated. Our results show the existence of the BG phase. In Section 4, we discuss several implications of the value of QCP from the viewpoint of $\textit{quantum percolation}$. Finally, Section 5 is devoted to the Conclusion.

\section{Model and numerical method}
\label{model and numerical method}
We study the KT transition in lattice Bose systems using eq.~(\ref{BH}) with the hard-core limit ($\textit{U}$$\rightarrow$$\infty$). This limit implies that the number $\hat {n}_i$ of particles at site $\textit{i}$ is restricted to 0 or 1. Then, eq.~(\ref{BH}) reduces to 
\begin{equation}
{\hat{\cal{H}}}=-J\sum_{<i,j>}(\hat b_i^{\dagger} \hat b_j+\hat b_i \hat b_j^{\dagger}) + \sum_i \mu_i \hat n_i.
\label{BH2}
\end{equation}
This model is equivalent to the $\textit{S}$=1/2 quantum XY model in a random field. We apply eq.~(\ref{BH2}) to a 2D square lattice of size $\textit{N}_s$=$\textit{L}^2$ with a periodic boundary condition. In this study, the random chemical potential $\mu_i$ is assumed to obey a Gaussian distribution given by
\begin{equation}
P(\mu_i)=\frac{1}{\sqrt{2\pi}\sigma}\exp\Big(-\frac{(\mu_i-\mu_0)^2}{2\sigma^2}\Big),
\label{gauss}
\end{equation}
where $\mu_0$ is the mean and $\sigma^2$ is the variance. In our calculations, $\textit{J}$ is fixed at 0.5 by the correspondence with the $\textit{S}$=1/2 quantum XY model\cite{Kawashima}, and the mean $\mu_0$ is $-1.0$. The effects of randomness are investigated by varying the variance $\sigma^2$.

The computational method used is path-integral QMC simulations using a modified directed-loop algorithm, which is one of the most efficient for calculating the lattice BH model\cite{Directed,Kawashima2,Kato}. We perform calculations for a maximum system size $L=250$ and maximum number of random sampling 512.

\section{Calculation and results}
\label{calculation and results}
In this section, we calculate the SF density $\rho_s$, the two-body correlation functions, and the compressibility $\kappa$. The main purpose is to investigate whether the KT transition occurs even in these disordered systems. The analysis procedure is the following. We perform the finite-size scaling of $\rho_s$ to determine precisely $T_{\rm KT}$. Then, we assess whether the correlation function below $T_{\rm KT}$ shows  a power-law decay. The variance $\sigma^2$ is varied from 0.0 to 8.8 to study the effects of randomness on the KT transition. We use the finite-size analysis in the same way as in the case of the pure KT transition in ref.\citen{Weber,Kawashima}. Even if the system is disordered, the symmetry of eq.~(\ref{BH2}) should be maintained, so that we can assume the finite-size analysis for the pure KT transition remains valid. In Section 3.4, we refer to the quantun critical phenomenon between the SF-BG phases in the present problem.

\subsection{SF density and estimation of the transition temperature}
\label{SF density and estimations of the transition temperatures}
The SF density $\rho_s$ is essential for studying the KT transition because it shows  the universal jump at $T_{\rm KT}$. It can be calculated directly by estimating the winding number $\boldmath{W}$ using path-integral QMC simulations\cite{SF}. The SF density $\rho_s$ is defined as
\begin{equation}
\rho_s=\frac{T}{2}\langle W^2 \rangle.
\label{SF}
\end{equation}

As an example, Fig.~1(a) shows the temperature dependence of the SF density $\rho_s(L)$ for various system sizes $L=24$, 32, 48, 64, 96, and 128 with $\sigma^2=1.5$, allowing us also to confirm the size dependence of the SF density. When the system size is relatively small, the SF density increases gradually as temperature decreases. As the system size increases, the SF density rises abruptly at a certain temperature. This seems to support the  universal jump at $T_{\rm KT}$ in this disordered system, if we extrapolate the behaviour to the case in which $\textit{L}$ is taken to infinity. 

What we want to know is the transition temperature $T_{\rm KT}$. The solid line in Fig.~1(a) represents $\rho_s=(2/\pi)$$\textit{T/J}$. The intersection between the solid line and $\rho_s(\infty)=\lim_{\textit{L}\rightarrow \infty}$$\rho_s(\textit{L})$ gives $T_{\rm KT}$ for the thermodynamic limit. The problem is estimating $\rho_s(\infty)$ from data on $\rho_s(\textit{L})$. As mentioned above, the results of numerical simulations are affected by finite-size effects. In order to obtain a reliable $T_{\rm KT}$, we follow the method of former studies\cite{Weber,Kawashima} to perform the finite-size analysis of $\rho_s(L)$ for this disordered system. According to these studies, the finite-size correction of $\rho_s$ at $T=T_{\rm KT}$ can be given by
\begin{equation}
\rho_s(L)=\frac{2 T_{\rm KT}}{\pi}\Big(1+\frac{1}{2 \log (L/L_0(T_{\rm KT}))}\Big),
\label{correction}
\end{equation}
where $\textit{L}_0$ is the characteristic length of the order of the lattice constant. It is notoriously difficult to determine $T_{\rm KT}$ precisely from this finite-size correction\cite{Kawashima}, however, we propose a simple method to obtain $T_{\rm KT}$ precisely by plotting $1/(\rho_s \pi/T - 2)$ as a function of $\textit{L}$. 

We rewrite eq.~(\ref{correction}) as 
\begin{equation}
\log L- \log L_0 (T_{\rm KT}) = \frac{1}{\rho_s(L) \pi/T_{\rm KT} -2}.
\label{correction2}
\end{equation}
According to eq.~(\ref{correction2}), we assume that $1/(\rho_s(\textit{L}) \pi/T_{\rm KT} - 2)$ should behave like $\log \textit{L} - \log \textit{L}_0 (T_{\rm KT})$ as a function of $\textit{L}$. Figure~1(b) is a rescaled plot of the SF density of Fig.~1(a) showing $1/(\rho_s \pi/T - 2)$ as a function of $\textit{L}$ for various temperatures. The bold curve in Fig.~1(b) represents $\log \textit{L} - \log \textit{L}_0 (T_{\rm KT})$.  Here, we regard $\textit{L}_0$ as a fitting parameter because changes of $\textit{L}_0$ only shift the plot of $\log \textit{L} - \log \textit{L}_0 (T_{\rm KT})$ versus $\textit{L}$ vertically without changing its shape. We vary $\textit{L}_0$ to obtain the value at which the plot of $\log \textit{L} - \log \textit{L}_0 (T_{\rm KT})$ fits the rescaled data of $1/(\rho_s \pi/T - 2)$. When $\textit{L}_0$ is 0.90($\pm$0.02), the plot of $\log \textit{L} - \log \textit{L}_0 (T_{\rm KT})$ fits $T/J=0.166(\pm0.02)$ (See Fig.~1(b)). Thus, the transition temperature $T_{\rm KT}$ with $\sigma^2=1.5$ is estimated to be 0.166($\pm$0.02)$\textit{J}$. This logarithmic dependence of $\rho_s(L)$ at exactly $T_{\rm KT}$ is the key to confirming the KT transition. We perform this finite-size scaling to determine $T_{\rm KT}(\sigma^2)$ for $\sigma^2$ from 0.0 to 8.8 and confirm the logarithmic behavior of the KT transition.
\begin{figure}[htbp]
% \begin{minipage}{0.5\hsize}
 \begin{center}
   (a)\includegraphics[width=70mm]{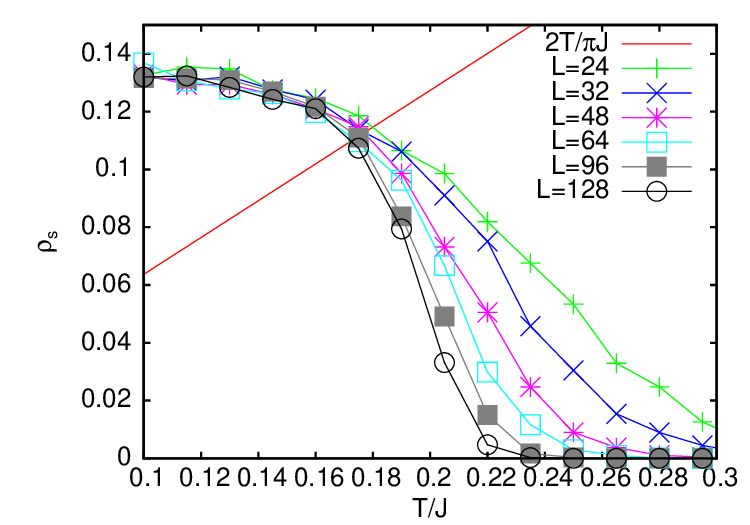}
  \end{center} 
%\caption{(a):(color online)Temperature dependence of the superfluid density with $\sigma^2$=0.0 for various system size. We can see the size dependence of the superfluid density. The dashed lines are a guide to the eye. The solid line is 2/$\pi$$\textit{T}$/$\textit{J}$.}
%  \label{fig:one}
% \end{minipage}
% \begin{minipage}{0.5\hsize}
  \begin{center}
   (b)\includegraphics[width=70mm]{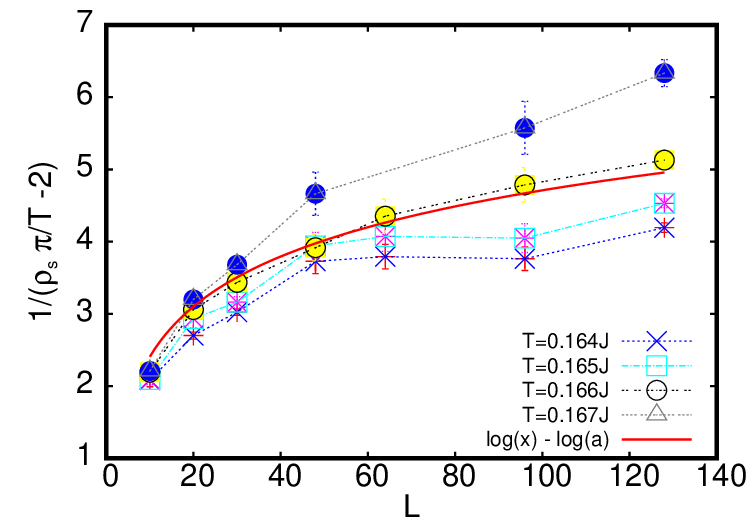}
  \end{center}
%\end{minipage}
  \caption{(a) (color online) Temperature dependence of the superfluid density for various system sizes with $\sigma^2=1.5$. The size dependence of the SF density is apparent. The lines connecting points are guides to the eye. The solid line is $(2/\pi)\textit{T}$/$\textit{J}$. (b) (color online) Rescaled plot of Fig.~1(a). The solid curve is the fit to $\log \textit{L} - \log \textit{L}_0 (T_{\rm KT})$.}
%  \label{fig:two}
% \end{minipage}
\end{figure}

\subsection{Correlation functions}
\label{Correlation functions}
An important characteristic of the KT transition is the existence of the QLRO showing the power-law decay below $T_{\rm KT}$. We calculate the two-body correlation function of a distance $\textit{r}$ to investigate the power-law below $T_{\rm KT}$, as follows:
\begin{equation}
G(r)=\frac{1}{N_s} \sum_i \langle \hat b_i ^\dagger \hat b_{i+r} \rangle.
\label{correlation}
\end{equation}
\begin{figure}[htbp]
\begin{center}
  \includegraphics[width=70mm]{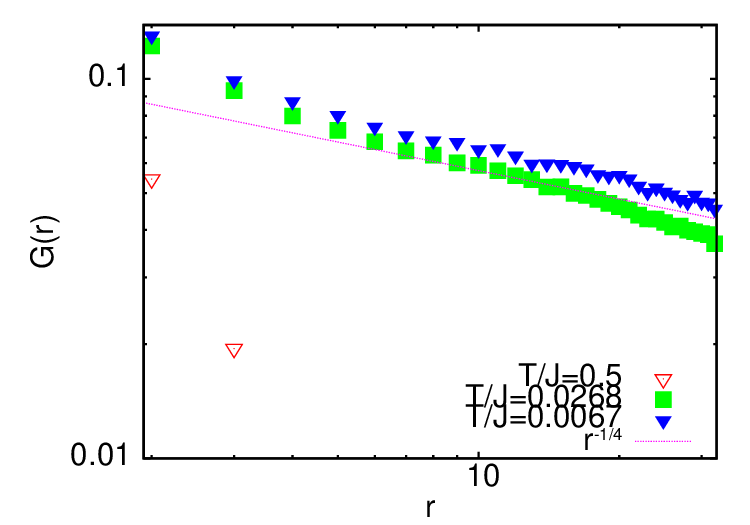}
\end{center}
% figure caption is below he figure
\caption{(color online) Correlation functions with $\sigma^2=6.0$ for various temperatures. We can observe the power-law decay of the correlation functions below $T_{\rm KT}$$\approx$0.0268$\textit{J}$.}
\end{figure}%
Figure~2 shows the correlation functions with $\sigma^2=6.0$ for three temperatures, namely $\textit{T}$=$T_{\rm KT}$($\approx $0.0268$\textit{J}$), $T>T_{\rm KT}$ (0.5$\textit{J}$), and $T<T_{\rm KT}$ (0.0067$\textit{J}$), for a finite system size ($L=128$). The critical exponent of the correlation functions $\eta(T_{\rm KT})$ is determined to be $\frac{1}{4}$ theoretically and numerically. It is clear that the numerical data supports $\eta(T_{\rm KT})$=$\frac{1}{4}$ from the solid line in Fig.~2 and the power-law decay below $T_{\rm KT}$. For each value of $\sigma^2$, the correlation function below $T_{\rm KT}$ decays according to the power law. This indicates the existence of the QLRO and also confirms the existence of the KT transition even in disordered systems.

\subsection{Phase diagram}
\label{phase diagram}
Figure~3(a) shows the phase diagram in the $\sigma^2$-$\textit{T}/\textit{J}$ plane. The points correspond to the numerical data of $T_{\rm KT}(\sigma^2)$ obtained by QMC simulations, and the solid line connecting the points is a guide to the eye. The transition temperature $T_{\rm KT}$ decreases as $\sigma^2$ increases, but the SF phase survives within a broad region of $\sigma^2$. This indicates that the SF phase is robust against randomness.

This robustness can be understood from the viewpoint of the Harris criterion\cite{Harris}. The Harris criterion is applied when considering the relevancy of randomness to the second-order phase transition. Near the critical temperature $T_c$, the specific heat $\textit{C}$ obeys $\textit{C}\sim|T_c/(T-T_c)|^\alpha$, and the critical exponent $\alpha$ is a key to the Harris criterion. According to the Harris criterion, randomness is relevant when the critical exponent $\alpha$ is positive. The properties of disordered systems qualitatively differ from those of pure systems. On the other hand, the randomness is irrelevant when the critical exponent $\alpha$ is negative (the specific heat does not diverge); the criticality does not change even in disordered systems. In the present case, since we study the KT transition, the specific heat does not diverge and shows a broad peak. Although the critical exponent $\alpha$ for the KT transition cannot be defined, randomness can be expected to be irrelevant to the KT transition if we consider the non-diverging specific heat.
\begin{figure}[htbp]
% \begin{minipage}{0.5\hsize}
  \begin{center}
   (a)\includegraphics[width=70mm]{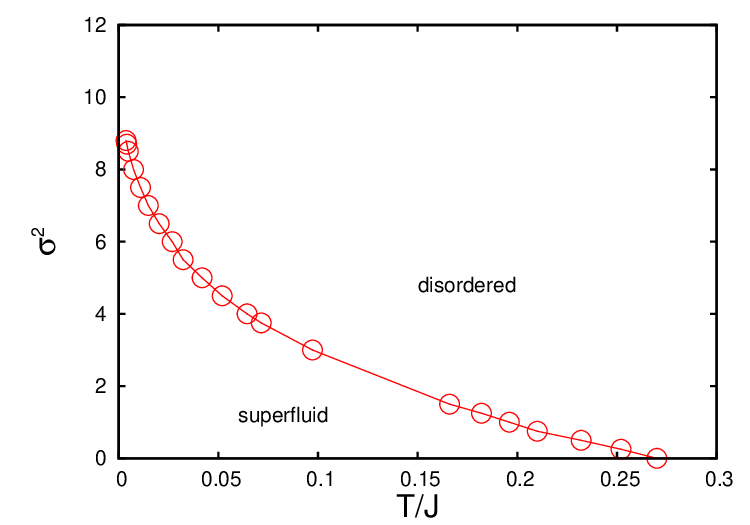}
  \end{center}
%  \caption{(a):The phase diagram in $\sigma^2$-$\textit{T}/\textit{J}$ plane.}
%  \label{fig:one}
% \end{minipage}
% \begin{minipage}{0.5\hsize}
  \begin{center}
   (b)\includegraphics[width=70mm]{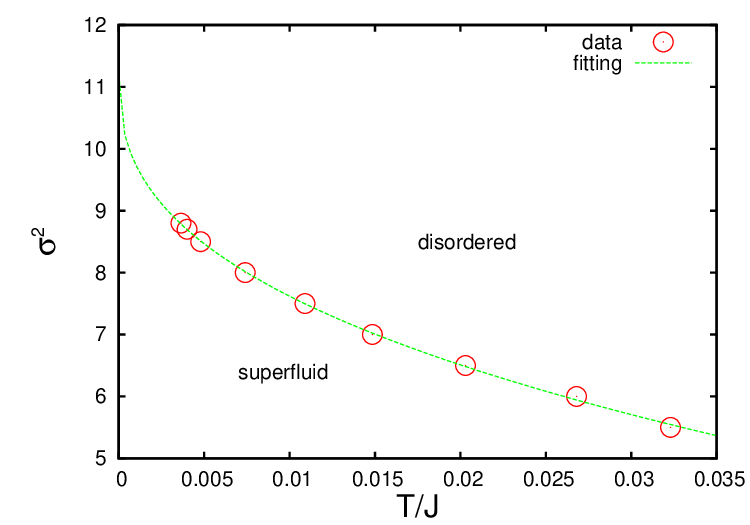}
  \end{center}
%\end{minipage}
  \caption{(color online) (a) Phase diagram in the $\sigma^2$-$\textit{T}/\textit{J}$ plane. The solid line is a guide to the eye. (b) A enlarged plot of Fig.~3 (a). From the fitting, $\sigma_c^2$ and $z\nu$ are estimated to be about 11.2 and about 2.57. The dotted curve represents the fitting.}
%  \label{fig:two}
% \end{minipage}
\end{figure}

\subsection{Quantum critical phenomenon between the SF-BG phases}
\label{quantum critical phenomenon}
Although we conclude the KT transition is robust against the randomness, $T_{\rm KT}$ goes to 0 at the QCP, $\sigma^2=\sigma_c^2$. Here, we focus on the quantum critical phenomenon in the present case. The system should enter the BG phase that is the localized-insulating state induced by the disorder appearing from the SF phase via the QCP. In this case, the parameter that controlls the disorder is $\sigma^2$. Thus, the correlation length along the imaginary time is described as $\xi_\tau \sim |\sigma^2 - \sigma_c^2|^{-\nu}$. The phase boundary in a finite temperature can be regarded as the cross-over line from the finite temperature to the zero temperature\cite{Cardy}. The line can be written as

\begin{equation}
T_{\rm KT}/J \sim|\sigma^2-\sigma_c^2|^{z\nu}
\end{equation}
In order to discuss the quantum critical phenomenon, we fit the data in the low-temperature region into the form $\sigma^2=A (T/J)^{1/z\nu} + \sigma_c^2$ ($A$ is a constant) (See Fig.3 (b)). The best fit gives us the parameters we need, $\sigma_c^2=11.2(\pm0.3)$ and $z\nu=2.57(\pm0.08)$. Then, we investigate the properties of the system for $\sigma^2>\sigma_c^2$.  We calculate the SF density and the compressibility. Figure~4(a) shows the temperature dependence of the SF density with $\sigma^2=0.0$, 2.0, 4.0, and 15.0 ($L=40$). As $\sigma^2$ increases, the SF density decreases. Clearly, the SF density with $\sigma^2=15.0$ vanishes. The compressibility $\kappa$ is defined as
\begin{equation}
\kappa=\frac{1}{N_s}\frac{\partial N}{\partial \mu}.
\label{compress}
\end{equation}
Here, $\textit{N}$ is the total number of particles. Figure~4(b) shows the temperature dependence of the compressibility. The data shown in Fig.~4(b) are for $\sigma^2=0.0$, 2.0, and 15.0, $\mu_0=-1.0$, and $L=40$. In addition, we plot the data of the insulating phase with $\sigma^2=0.0$, $\mu_0=-5.0$, and $L=40$. The compressibility is calculated for various system sizes, but little size dependency of compressibility is apparent. In the insulating phase, for the lower temperature, the compressibility vanishes. As the temperature increases, the compressibility has an thermal-activated form $e^{-\triangle_{gap}/T}$ with a finite energy gap $\triangle_{gap}$\cite{Min}. With $\mu_0=-1.0$, when $\sigma^2$ is 0.0, we can see that the compressibility has a sharp peak. The peak becomes broad as $\sigma^2$ increases, but the compressibility remains finite. With $\sigma^2=15.0$, although the system seems to be in the insulating phase because of the zero SF density, the compressibility remains finite even if the temperature becomes zero, just as in the case $\sigma^2<\sigma_c^2$. This result shows the existence of the BG phase in this disordered system. The emphasis in this papaer is that we obtain the precise estimation of $z\nu$ by means of QMC simulations rather than the former studies\cite{MC1,MC2,dual,real}. The dynamical critical exponent $z$ of the SF-BG transition is estimated to be 2.0\cite{MC1,2D} so that $\nu$ can be estimated 1.28. This value is consistent with $\nu=1.15(\pm0.1)$\cite{MC2} within errorbars.
\begin{figure}[htbp]
% \begin{minipage}{0.5\hsize}
  \begin{center}
   (a)\includegraphics[width=73mm]{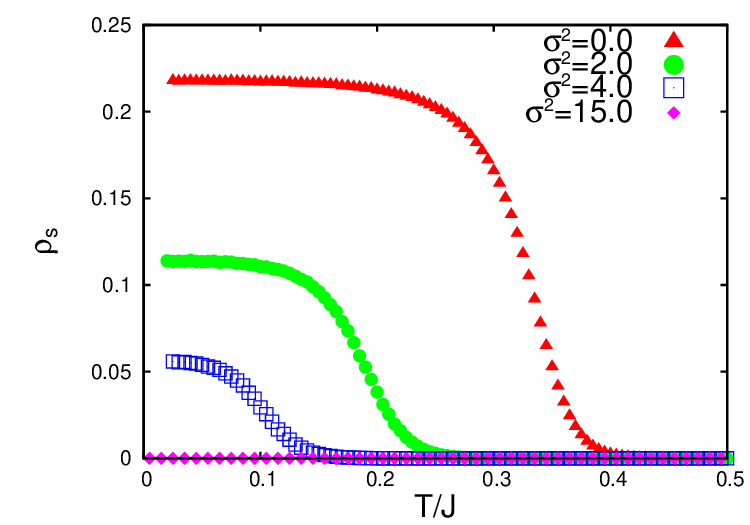}
  \end{center}
%  \caption{(a): Phase diagram in the $\sigma^2$-$\textit{T}/\textit{J}$ plane.}
%  \label{fig:one}
% \end{minipage}
% \begin{minipage}{0.5\hsize}
  \begin{center}
   (b)\includegraphics[width=73mm]{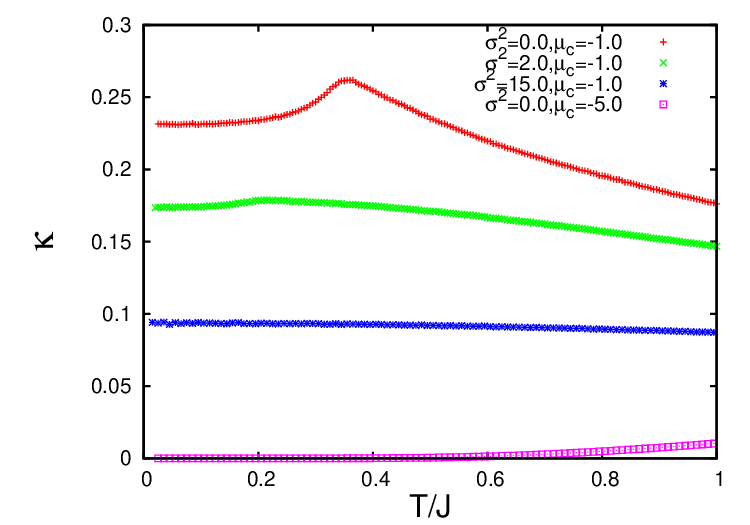}
  \end{center}
%\end{minipage}
  \caption{(color online) (a) Temperature dependence of the SF density for various variances with $L=40$. (b) Temperature dependence of compressibility with $L=40$.}
%  \label{fig:two}
% \end{minipage}
\end{figure}

\section{Discussion}
\label{Discussion}
In this section, we consider what the computed value of $\sigma_c^2$ indicates. We consider the phase diagram of the ground state in a pure system ($\sigma^2$=0.0 and $\mu_i$=$\mu$ in eqs.~(\ref{BH}) and~(\ref{BH2})) analytically and numerically to study the relation between the SF phase and the chemical potential. This relation leads us to compare the present problem with the 2D site-percolation problem.

$\textit{Analytical approach}$: In our calculations, the hard-core limit was imposed on eq.~(\ref{BH}). This limit implies that the number of particles at each site is limited to 0 or 1. The phase boundary between the insulator and the SF phases for the pure system can be obtained easily under this restriction. For large $\mu$, the ground state is the insulator phase given by $|\psi_0\rangle=|1,1,\dots,1\rangle=\prod_{i=1}^{N_s} |1\rangle_i$, in which a single particle occupies each site. The first excited state given by $|\psi_1 \rangle=\frac{1}{\sqrt{N_s}} \sum_{k=1}^{N_s}|a_k \rangle$ is the superposition of $|a_k\rangle$, where $|a_k\rangle$ is the state in which a site $\textit{k}$ is vacant while other sites ($\textit{i}\neq\textit{k}$) are occupied. Comparing the energy between the ground and the first excited state, we can obtain the phase boundary of $|\mu|=4J$ in the 2D system.

$\textit{Numerical approach}$: We investigate the chemical potential dependence of the average of the number of particles at each site $\textit{n}$=$\frac{N}{N_s}$.
%\begin{equation}
%n=\frac{\rho}{N}.
%\end{equation}
Figure~5 shows an example for hopping parameter $J=0.5$ at various temperatures. As mentioned above, the transition point $\mu_c$ between the insulator and the SF phases is defined as the point at which $\textit{n}$ changes from unity. The temperature dependence of $\mu_c$ is apparent from the figure. By extrapolating $\mu_c(T)$ as a function $\textit{T}$, $\mu_c(0)$ can be determined.

\begin{figure}[htbp]
\begin{center}
  \includegraphics[width=70mm]{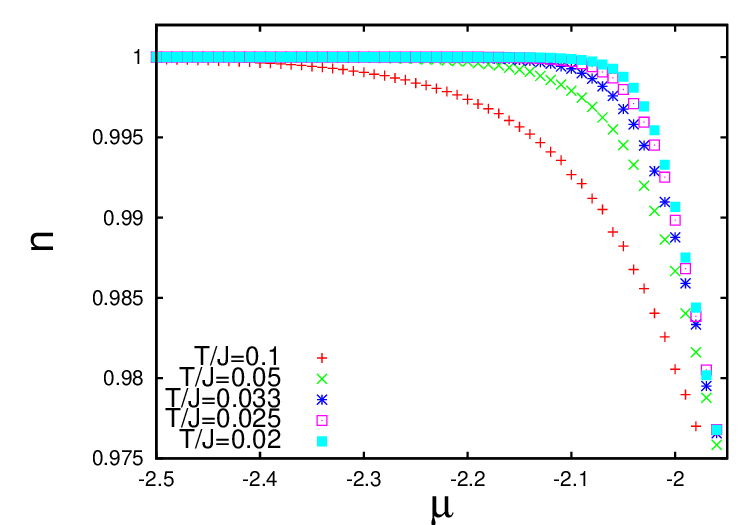}
\end{center}
% figure caption is below he figure
\caption{(color online) Chemical potential dependence of the average number of particles at each site for $J=0.5$ at various temperatures.}
\label{fig:2}       % Give a unique label
\end{figure}%

%\end{figure}%

In this way, we obtain the phase diagram of the ground state in the pure system (Fig.~6). The dotted lines are the analytical results $|\mu|=4\textit{J}$, and the points are the numerical results. Clearly there is a nice agreement between these results. The central region is the SF phase, and the upper and lower regions are the insulator phases.
\begin{figure}[htbp]
\begin{center}
  \includegraphics[width=70mm]{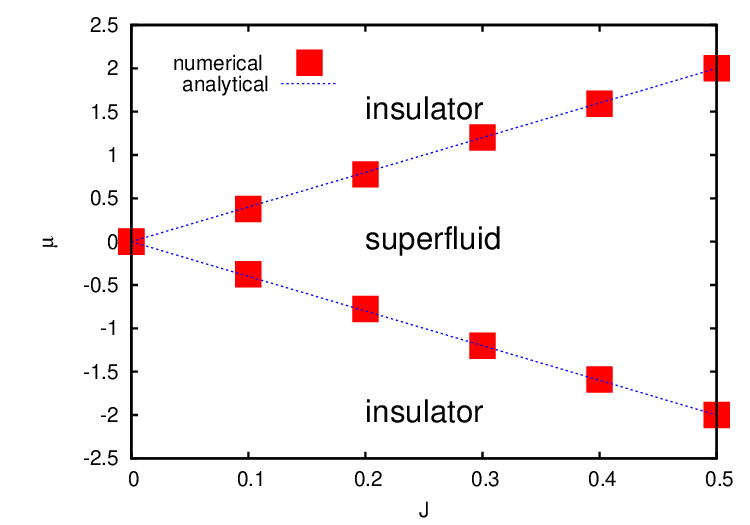}
\end{center}
% figure caption is below he figure
\caption{(color online) Phase diagram of the ground state in a 2D pure system. The dotted lines are the analytical results and the points are the numerical results. The central region is the superfluid phase, and the upper and lower regions are the insulator phases. }
\end{figure}%

In our study, the SF state can be described as the state that the correlation of particles develops over the whole system, namely, a particle moves over the whole system during the imaginary time development. We consider what the value of $\sigma_c^2$ indicates from the viewpoint of percolation by comparing Fig.~6 and QMC simulations for our disordered case at $T=0$. In QMC simulations for the disordered case, the random distribution of the chemical potential was assumed to be the Gaussian distribution. The mean $\mu_0$ was $-1.0$ and the variance $\sigma_c^2$ was calculated to be 11.2. The hopping parameter $\textit{J}$ was fixed at 0.5. As shown in Fig.~6, for the pure system, the system shows superfluidity in the region of $-2 \leq \mu \leq 2$ at $J=0.5$. The chemical potential $\mu=\pm$2 can be considered as the critical chemical potential  for systems to show superfluidity. Next, the probability $P_{\rm G}$ that the chemical potential at a site takes a value within $-2 \leq \mu \leq 2$ when the random distribution of the chemical potential obeys the Gaussian distribution ($\mu_0=-1.0$, $\sigma_c^2=11.2$) is calculated to be about 43$\%$. If the probability that the chemical potential takes a value within $-2 \leq \mu \leq 2$ exceeds $P_{\rm G}$, the system can enter the SF phase. This probability $P_{\rm G}$ should be considered as the threshold for entering the SF phase.

The values of probability $P_{\rm G}$ reminds us of the 2D site-percolation problem. It is the geometrical connection that is important for the site-percolation problem. In this picture, the sites exist randomly. That is, we assign only two energy levels, zero or infinity, randomly to the site energy. When the probability that the zero-energy sites are distributed exceeds the 2D site-percolation threshold $P_{\rm c}$, which was obtained to be 59.3$\%$ numerically\cite{percolation}, a spanning cluster exists and the system becomes coherent. If we map our disordered case to this two-level system, sites such that -2$\leq \mu \leq$2 might be considered as the zero-energy sites and the others as the infinite-energy sites. However, no precise correspondence between $P_{\rm G}$ and $P_{\rm c}$ can be seen. Apparently, $P_{\rm G}$ is about 16$\%$ smaller than $P_{\rm c}$. Therefore, we wish to know what is underlying this difference. 

There are two reasons why $P_{\rm G}$ and $P_{\rm c}$ should be different. The first reason is the difference of the percolation process. If the percolation process is regarded as dynamical, meaning that particles propagate infinitely far from their initial position, whether the dynamics of the particle is classical or quantum makes a clear difference to the percolation process. The threshold $P_{\rm c}=59.3\%$ was obtained under the classical picture. In this picture, the geometrical connection can be understood as the propagation of the particles through zero-potential sites. On the other hand, in the quantum percolation description, not only the site energy but also the state of the site, namely, the number of particles $\hat n_i$ for each site, must be considered. Our numerical results for the disordered systems refer to the limit of the dynamical percolation process along imaginary time as $\beta$(=1/$T$)$\rightarrow$$\infty$, and the state of the site is the number of particles. Thus, our case should be dominated by the quantum percolation process. The second reason is that the site energy is not infinite but has a finite value in the present case. Under the quantum picture, the infinite site energy blocks propagation of the quantum particle, causing a quantum localization effect. Because of this localization effect, the quantum percolation threshold is larger than $P_{\rm c}$ in this case\cite{odagaki}. However, it is known that the finite site potential reduces the quantum percolation threshold below $P_{\rm c}$. If the site energy is finite, the quantum particle can tunnel through the potential barriers, and it tends to propagate infinitely far from the initial position, unlike in the classical description\cite{onogi}. In our case, the site energy takes a finite value that obeys a Gaussian distribution. Together, these two reasons explain why $P_{\rm G}$ should be expected to be smaller than $P_{\rm c}$.

\section{Conclusion}
\label{Conclusion}
In this paper, we described, using QMC simulations, a study of the 2D hard-core BH model with a random chemical potential that was assumed to obey a Gaussian distribution. The KT transition was confirmed from a finite-size analysis of the SF density $\rho_s$ and the power-law decay of the correlation function. In order to investigate the effects of randomness on the KT transition, the variance of the distribution was varied, and we obtained a phase diagram showing the SF and the disordered phases. Even if the variance increases, the KT transition survives robustly against the randomness. The Harris criterion helps to explain this robustness of the KT transition. We also investigated the quantum critical phenomenonon of the SF-BG transition. The parameters, $\sigma_c^2$ and $z\nu$, are precisely estimated to be 11.2 and 2.57, respectively.

In addition, we considered what this value of $\sigma_c^2$ indicates and argue that it can be understood from the viewpoint of percolation. Comparing the 2D percolation threshold and the present case makes the quantum effects in the percolation process clear. Our results should be dominated by the quantum percolation processes.

\begin{acknowledgements}
Part of the calculation was done using the facilities of the Supercomputer Center at the Institute for Solid State Physics, University of Tokyo. M. Tsubota acknowledges the support of a Grant-in-Aid for Scientific Research from JSPS (Grant No. 21340104).
\end{acknowledgements}

\end{document}